\documentclass[aps,twocolumn,superscriptaddress]{revtex4}
\usepackage{epsfig,dsfont,amssymb,amsmath,amsthm,amsfonts,amsbsy,mathrsfs}
\usepackage{epsfig,amssymb,amsmath,amsthm,amsfonts,amsbsy,mathrsfs}
\usepackage{graphicx}
\usepackage{color}
\usepackage{bm}

\begin{document}
\title{Reverse Janssen effect in narrow granular columns}

\author{Shivam Mahajan}
\affiliation{Division of Physics and Applied Physics, School of Physical and
Mathematical Sciences, Nanyang Technological University, Singapore}

\author{Michael Tennenbaum}
\affiliation{School of Physics, Georgia Institute of Technology, Atlanta, GA 30332, USA}

\author{Sudhir N. Pathak}
\affiliation{Division of Physics and Applied Physics, School of Physical and
Mathematical Sciences, Nanyang Technological University, Singapore}

\author{Devontae Baxter}
\affiliation{School of Physics, Georgia Institute of Technology, Atlanta, GA 30332, USA}

\author{Xiaochen Fan}
\affiliation{School of Physics, Georgia Institute of Technology, Atlanta, GA 30332, USA}

\author{Pablo Padilla}
\affiliation{School of Physics, Georgia Institute of Technology, Atlanta, GA 30332, USA}

\author{Caleb Anderson}
\affiliation{School of Physics, Georgia Institute of Technology, Atlanta, GA 30332, USA}

\author{Alberto Fernandez-Nieves}
\email{alberto.fernandez@physics.gatech.edu}
\affiliation{School of Physics, Georgia Institute of Technology, Atlanta, GA 30332, USA}
\affiliation{Department of Condensed Matter Physics, University of Barcelona, 08028 Barcelona, Spain}
\affiliation{ICREA-Institució Catalana de Recerca i Estudis Avançats, 08010 Barcelona, Spain}

\author{Massimo Pica Ciamarra}
\email{massimo@.ntu.edu.sg}
\affiliation{Division of Physics and Applied Physics, School of Physical and
Mathematical Sciences, Nanyang Technological University, Singapore}
\affiliation{CNR--SPIN, Dipartimento di Scienze Fisiche,
Universit\`a di Napoli Federico II, I-80126, Napoli, Italy}

\date{\today}

\begin{abstract}
When grains are added to a cylinder, the weight at the bottom is smaller than the total weight of the column, which is partially supported by the lateral walls through wall/grain frictional forces. This is known as the Janssen effect.
Via a combined experimental and numerical investigation, here we demonstrate a reverse Jansen effect whereby the fraction of the weight supported by the base overcomes one.
We characterize the dependence of this phenomenon on the various control parameters involved, rationalize the physical process responsible for the emergence of the compressional frictional forces responsible for the anomaly, and introduce a model to reproduce our findings.
Contrary to prior assumptions, our results demonstrate that the constitutive relation on a material element can depend on the applied stress.
\end{abstract}

\maketitle

Driven by the need of designing strong enough silos to contain granular particles, Janssen~\cite{Janssen,Sperl,RevJaeger,Andreotti} investigated the forces that grains exert on their container.
He observed that the force $f$ exerted on the base of the container, or equivalently the apparent mass $m_a(h) = f/g$, with $g$ the gravitational acceleration, saturated on increasing the filling height $h$.
This indicates that an increasingly larger fraction of the added grain mass is supported by the walls of the container through frictional forces.
This experiment later emerged as a standard benchmark for the theories of granular elasticity. 
Janssen himself developed a continuum phenomenological model. 
He assumed the radial component of the stress to be constant and proportional to the longitudinal component through a factor $k$, and that the contacts between the particles and the walls were at their Coulomb threshold.
This is essentially the assumption of the Incipient Failure Everywhere approach~\cite{Nedderman1992}.
With these assumptions the apparent mass is found to saturate exponentially with the added mass, $m(h)$: $m_a(h) = m_\infty\left[1-\exp(-m(h)/m_\infty)\right]$, where $m_\infty/m(h)=\lambda/h$ and $\lambda = D/(4 \mu k)$, with $D$ the diameter of the cylindrical silo and $\mu$ Coulomb's friction coefficient.
This theoretical prediction has been experimentally investigated considering protocols able to bring the particle wall contacts to the Coulomb limit~\cite{Sperl,VanelEPJB,VanelPRL,BratbergUpward}, as assumed by Janssen.
In this limit, the one-parameter, $k$, Janssen model reasonably describes the experimental results, only slightly underestimating the apparent mass at small filling heights, and slightly overestimating it at large ones~\cite{VanelEPJB,VanelPRL,Shaxby1923}.
A two-parameter model derived within the Fixed Principal Stress Axes (FPSA) or the Oriented Stress Linearity (OSL) model~\cite{Nedderman1992,Wittmer1997,VanelPRL,BratbergUpward}, provides a better description of the experimental results.

Janssen's assumptions are generally not met when a silo is simply filled by pouring grains into it, and indeed a dependence of $m_a(h)$ on the filling protocol was found in early experiments~\cite{Shaxby1923,JotakiDependFillingProcedure}.
Shaxby~\cite{Shaxby1923} suggested this dependence results from different protocols giving rise to different `surfaces of equal pressure', somehow anticipating the OSL model~\cite{Nedderman1992,Wittmer1997,VanelPRL,WittmerLetterstoNature}.
In Janssen's model, these surfaces are horizontal, as the radial component of the stress is proportional to the longitudinal component.
On the contrary, Shaxby noticed that these surfaces could actually be convex or concave, depending on whether the frictional forces at the wall support the grains, or rather compress the grains.
However, frictional forces that compress the grains have never been reported so far.

In this Letter, we show, via a combined numerical and experimental investigation, that the simplest protocol one could devise to fill a cylinder with grains, their sequential deposition, leads to the emergence of frictional compressive forces on the grains. 
These forces are always present, but only occur close to the free surface of the granular column. 
In shallow cylinders, $h \leq 30 \sigma$, with $\sigma$ the diameter of the grains, these compressive forces play a dominant role, and lead to a reverse Janssen effect whereby the apparent mass overcomes the true mass. 
This reverse Janssen effect is apparent in small cylinders and progressively disappears as $D$ increases, as we rationalize introducing a model that correctly reproduces our findings.

In the experiments, we consider spherical plastic beads of diameter and mass $\sigma = (5.94 \pm 0.02) \textrm{mm}$ and $M = (112.6 \pm 0.1)$mg, respectively, and glass cylinders with different $D$. 
The cylinders are held above a scale, and the distance between the lower edge of the cylinder and the scale is much smaller than the grain diameter.
Grains are inserted in the cylinder in small chunks, and the system reaches mechanical equilibrium before new grains are added.
This process is repeated until the column is filled up to a desired height and the apparent mass and column height are recorded during the process.

\begin{figure}[!t]
 \centering
 \includegraphics[angle=0,width=0.48\textwidth]{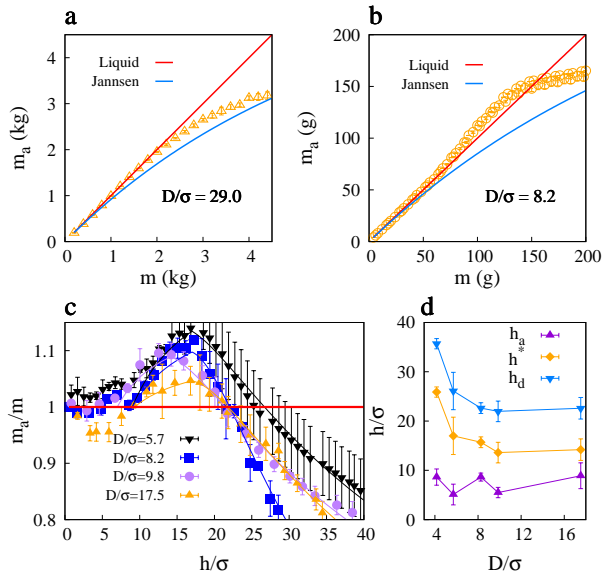}
 \caption{
  Experimental results for the dependence of the apparent mass, $m_a$, on the added mass, $m$, for a cylinder with diameter $D=(172.00 \pm 0.15)$mm $\simeq 29 \sigma$, panel a, and $D=(49.10 \pm 0.15)$mm $\simeq 8.2 \sigma$, panel b, with $\sigma$ the grain diameter.
Error bars are standard deviations from the average over $3$ different realizations. 
    c. Dependence of the ratio of the apparent to added mass, $m_a/m$, on the filling height relative to the grain diameter, $h/\sigma$, for different $D$. Full lines are fit to the theoretical model, with 
    $\xi \simeq 5.0\sigma$, $h_a \simeq 9\sigma$, $h^* \simeq 17\sigma$, $F_c=-1.4$Mg, and $k$ varying in the range $[0.03,0.12]$.
d. Characteristic length scales corresponding to ($h_a$) the height where the deviation from the hydrodynamic expectation starts, ($h^*$) the height corresponding to the maximum in $m_a/m$, and ($h_d$) the height to the right of the maximum where $m_a/m = 1$.
  \label{fig:exptAnom}
}
\end{figure}
Fig.~\ref{fig:exptAnom}a illustrates the dependence of the apparent mass on the added mass for a cylinder with diameter $D = (172.00 \pm 0.15) \textrm{mm} \simeq 29\sigma$. 
For this large cylinder we observe a Janssen like behavior, which is only approximately described by Janssen model as our protocol does not assure that the particle-wall contacts are at their Coulomb threshold. 
In panel b, we show the same plot for a smaller cylinder, $D = (49.10 \pm 0.15) \textrm{mm} \simeq 8.2\sigma$.
The striking feature here is the existence of an anomalous regime where the experimental data lie above the $m_a = m$ line characterizing the behavior of liquids, further indicating that $m_a > m$ for certain filling heights.
Beyond this anomalous regime $m_a$ saturates.
We have explicitly checked that this anomalous effect does not disappear if we gently tap the cylinder every time we add grains.
This anomalous behavior is captured neither by Janssen's model, nor by its two-parameter generalizations~\cite{Shaxby1923,VanelPRL}.
In the following, we first experimentally and numerically investigate this anomalous effect, and then develop a model accounting for the observed behavior.

Fig.~\ref{fig:exptAnom}c illustrates the dependence of the normalized apparent mass $m_a/m$ on the normalized filling height, $h/\sigma$, i.e. the number of layers, for cylinders with different diameters.
We observe that the apparent mass $m_a$ overcomes the added one $m$ at a filling height $h_a$, and becomes smaller than it for $ h > h_d$. The maximum in $m_a/m$ occurs at $h=h^*$. 
These characteristic length scales are essentially independent on the cylinder diameter, as shown in panel d, if not for very small $D$.
The existence of a filling height $h_a$ below which our granular columns follow the hydrodynamic expectation is consistent with previous observations~\cite{VanelEPJB}, and is rationalized considering that there are essentially no frictional interactions with the wall in very shallow columns.
Fig.~\ref{fig:exptAnom}c also indicates that the strength of the anomaly, as quantified by the maximum value of $m_a/m$, decreases on increasing the cylinder diameter.
Since deviations from the hydrodynamic expectation $m_a = m$ can only originate from frictional forces at the wall, the observed behavior indicates there are frictional forces that push the grains downwards. 
These compressive frictional forces seem to dominate the behavior in an intermediate range of filling heights, resulting in the observed anomaly.

To unveil the physical mechanism leading to these compressive forces, we perform molecular dynamics simulations~\cite{Plimpton1995a}.
We use monodisperse spherical particles of mass $M$ and diameter $\sigma$, as in the experiments, and add them sequentially to a cylinder of diameter $D$, where they settle under the influence of the gravitational acceleration. 
Specifically, we add a new particle only after the system reaches a state of mechanical equilibrium. 
We mimic the experimental gentle deposition process by inserting a new particle just above the deposited ones, with a random horizontal position within the cylinder. 
We use standard models for the normal and frictional interactions between the particles and between the particles and the wall~\cite{Silbert,SM}.
For each contact, we enforce Coulomb's law so that the magnitude of the tangential force $f_t$ satisfies $f_t \leq \mu f_n$, with $f_n$ the normal force.
We further enforce here $\mu = 0.2$, describing the role of the friction coefficient in the Supplemental Material~\cite{SM}.

Fig.~\ref{fig:anomalyAndLength}a illustrates numerical results for the dependence of $m_a/m$ on $h/\sigma$.  
The simulations qualitatively reproduce all the experimentally observed features.
In particular, the anomaly is well reproduced and occurs within a range of filling heights $h_a < h < h_d$, which does not depend on $D$, as shown in Fig.~\ref{fig:anomalyAndLength}b.
In the simulations, we have access to the forces acting between the particles and between the particles and the cylinder wall. 
In Fig.~\ref{fig:anomalyAndLength}c, we illustrate the largest forces present in the system with $D = 8 \sigma$ at a filling height $h = 10.2 \sigma$ at which we observe the anomaly in $m_a/m$. 
Large forces clearly radiate from the cylinder wall 
towards the bottom. 
These are the forces responsible for having a larger apparent mass compared to the added mass; see also Fig.~S1~\cite{SM}.
In contrast, in Fig.~\ref{fig:anomalyAndLength}d we consider a larger filling height ($h=25\sigma$), where $m_a/m < 1$, and illustrate the frictional forces occurring deep in the column ($h < 13\sigma$). 
These forces majoritarily point from the cylinder wall to the top, hence contributing to sustaining the grains, as in the prototypical Janssen scenario. 

\begin{figure}[!t]
 \centering
 \includegraphics[angle=0,width=0.48\textwidth]{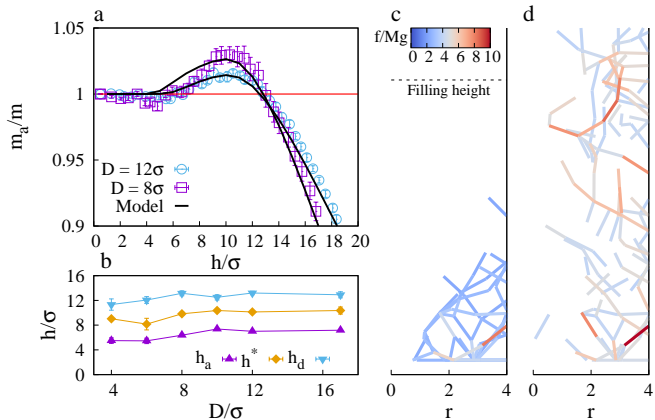}
 \caption{
 a. Ratio of apparent to total mass as a function of the normalized filling height, $h/\sigma$, for cylinders with different diameters. Error bars are standard deviations from the average over $3$ different realizations. 
 Full lines are fit to the theoretical model, with $\xi \simeq 5.6\sigma$, $h_a \simeq 5.6\sigma$, $h^* \simeq 10.5\sigma$, $F_c=-0.36$Mg and $k\simeq0.14$.
 b. Cylinder-diameter dependence of the characteristic heights where the reverse Janssen effect appears ($h_a$), reaches its maximum ($h^*$), and disappears ($h_d$). 
 c, d. Simulation snapshots illustrating the force network in the granular column with $D = 8 \sigma$. Each segment connects the positions $(r,h)$ of two interacting grains, with $r$ the radial position. The color indicates the magnitude of the interaction force according to the color-scale in c. Only the largest (5\%) forces are shown in each panel. In c, the filling height is in the anomalous region, while in d the filling height ($h = 25 \sigma$) is in the saturation region. In the latter case, we only show the forces deep in the column ($h < 13 \sigma$).
\label{fig:anomalyAndLength}
}
\end{figure}

\begin{figure}[!t]
 \centering
 \includegraphics[angle=0,width=0.48\textwidth]{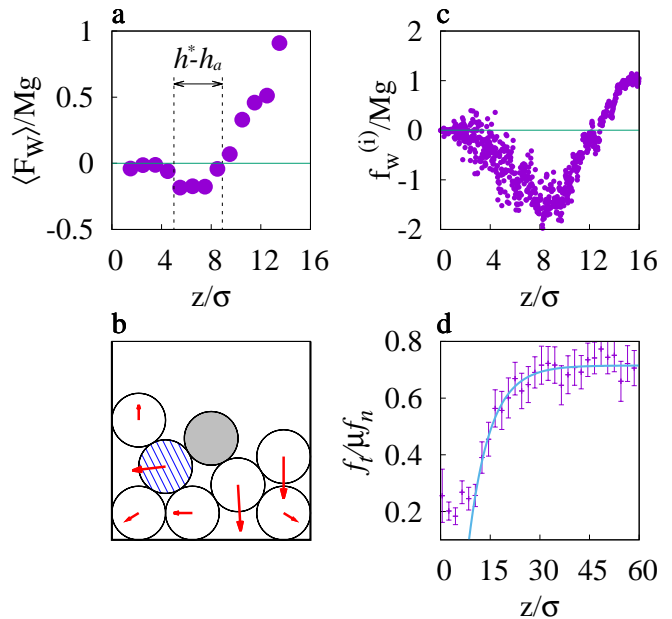}
 \caption{
 a. Average frictional force exerted by the wall on a particle as a function of the depth $z/\sigma$. Negative forces are compressive. 
 b. Numerical results for the displacements (red-arrows) of a stable frictionless packing of disks induced by the addition of a new grain, the shaded one.
 Shown displacement vectors have an increased length relative to the actual value of $10^5$.
 c. Force exerted by the wall on a given particle, as a function of its depth. The depth increases as more particles are added to the column. 
 d. Depth dependence of the average value of the ratio between the normal and tangential forces the wall exert on each particle. The ratio exponentially approaches a constant value, with length scale $\xi \simeq 6\sigma$.
 In panels a,c and d, the cylinder diameter is $D=8 \sigma$, while the friction coefficient is $\mu=0.2$.
\label{fig:forceWallInset}
}
\end{figure}

Compressive forces are present in the system regardless of the filling height.
This is demonstrated in Fig.~\ref{fig:forceWallInset}a where we investigate the dependence of the average wall/particle frictional force as a function of the particle {\it depth} $z$, for a cylinder with $D=8\sigma$ and $h= 25 \sigma$. Negative forces correspond to compressive forces.
The average force is $\langle F_w \rangle \simeq 0$ at small depths,
compressive in a subsequent depth range of approximate length $\sim h^*-h_a$, and supportive at larger depths,
The depth values for which forces first become compressive, and then supportive,
roughly correspond to those of $h_a$ and $h^*$, respectively.
The existence of compressive forces close to the top surface of the granular column clarifies that these forces are not induced by the bottom substrate.

To further rationalize the physical processes leading to the compressive forces, we recall that the frictional force acting between two objects depends on how much they are sheared one with respect to the other.
For instance, in the popular Mindlin model~\cite{49Min}, which is the one used in our numerical simulations, the frictional force is $-k_t {\bf s}$, with $k_t$ a tangential stiffness, and ${\bf s}$ the shear displacement.
In the case of a particle in contact with a fixed wall, the shear displacement is the integral of the particle velocity at the particle-wall contact point over the duration of the contact.
The compressive forces might therefore arise through a simple mechanical process inducing an upward particle motion, as we illustrate in Fig.~\ref{fig:forceWallInset}b, where we add a new disk, which is shaded in the figure, to a stable packing of frictionless disks. 
This causes a rearrangement, as the grains need to reach a new equilibrium configuration, which we illustrate by associating to each grain an arrow whose length is proportional to its displacement.
Clearly, the addition of a particle can induce the upward motion of particles in contact with the wall, as in the case of the leftmost top particle.
In the presence of friction, this upward movement would be counteracted by the wall-particle frictional interaction leading to the emergence of compressive frictional forces.
In addition, the possible rotation of the particle as it moves upwards, could equally lead to compressive wall-particle frictional forces~\cite{SM}.
The same scenario occurs in three dimensions. 
Notice that according to this picture no anomaly can occur when $D<2 \sigma$.
We do have explicitly checked this in our numerical simulations, see Fig.~S5~\cite{SM}.
In addition, we note that the upward movement is induced by the radial displacement of other particles, such as the striped one in Fig.~\ref{fig:forceWallInset}b, which acts as a wedge.
This radial displacement is reduced in the presence of frictional interactions, consistent with what happens in granular piles~\cite{Hentschel}.
Hence, we expect the anomaly to disappear in the high-friction limit. 
We have confirmed that the anomaly disappears in this case, and also in the low-friction limit, due to the weakness of the frictional forces~\cite{SM}.

The found dependence of the apparent mass on the filling height is captured neither by hydrodynamics nor by Janssen's model.
The FPSA model~\cite{Wittmer1997,VanelPRL} is also unable to describe our findings.
Indeed, the key assumption in the FPSA model is that the stresses on a `material element are fixed at the time of its burial, and unaffected by loads applied subsequently'~\cite{Wittmer1997}. This hypothesis is not fulfilled in our simulations.
We explicitly illustrate this is in Fig.~\ref{fig:forceWallInset}c, where the frictional force $f_w^{(i)}$ the wall exerts on a given particle $i$ is plotted as a function of the particle depth. 
The force changes with increasing load and depth.
Analogous results are found for other particles.

We rationalize our results starting from the continuity equation for the dependence of pressure $\Sigma$ on depth,
\begin{equation}
\frac {d\Sigma}{dz} = \rho g - \frac {4}{D} \tau(z),
\label{eq:Janssen}
\end{equation}
where $\tau(z)$ is the stress exerted by the wall on the particles. 
This equation can be solved by separation of variables if $\tau$ only depends on $\Sigma(z)$, e.g. as in Janssen's model. 
Since in our case, there are more relevant length scales than just $D$, this assumption does not hold; a similar situation was previously encountered with colloidal gels~\cite{Condre2007}, whose pressure profile was described by extending Janssen's model through the incorporation of an additional length scale related to the elasticity of the gel.
Here we consider that these additional length scales have a frictional origin, and we incorporate them in the depth dependence of the stress $\tau(z)$.
We take $\tau(z) = n_c(D) \langle F_w(z) \rangle /(\pi D)$, 
where $\langle F_w(z) \rangle$ is the average frictional force at depth $z$ per particle and $n_c(D)$ is the number of contacts per unit length of the cylinder~\cite{SM}. 
Fig.~\ref{fig:forceWallInset}a suggests that one may assume $\langle F_w(z) \rangle \simeq 0$ for $z < h_a$, and $\langle F_w(z) \rangle = F_c < 0$ for $h_a < z < h^*$. 
As a grain experiencing a compressive frictional force is buried through the addition of more grains, its depth increases.
The grain is also pushed downwards by the newly added grains, so that its interaction with the wall, which is initially compressive, changes to supportive as its depth increases.
Hence, deep in the granular column the frictional forces with the wall are supportive. 
Assuming that these forces will be at their Coulomb threshold, as in Jannesen's model, we postulate 
$\langle F_w(z) \rangle = k \pi D\Sigma(z)\left(1-e^{-(z-h^*)/\xi}\right)/n_c(D)$ for $z > h^*$. 
This assumption is supported by Fig.~\ref{fig:forceWallInset}d, where we show that the ratio $f_t^{(i)}/\mu f_n^{(i)}$ between tangential and normal forces at the wall, averaged over all particle-wall contacts at a given depth $z$, does approach a constant value exponentially, with a decay length $\xi \simeq 6 \sigma$.

The model correctly describes the numerical results for different $D$, without the need of adjusting any of its parameters, as 
shown in Fig.~\ref{fig:anomalyAndLength}a.
In the experimental case, we need to slightly change $k$ with $D$ to describe the data. Representative fits are shown in Fig.~\ref{fig:exptAnom}c.

\begin{figure}[tb]
 \centering
 \includegraphics[angle=0,width=0.4\textwidth]{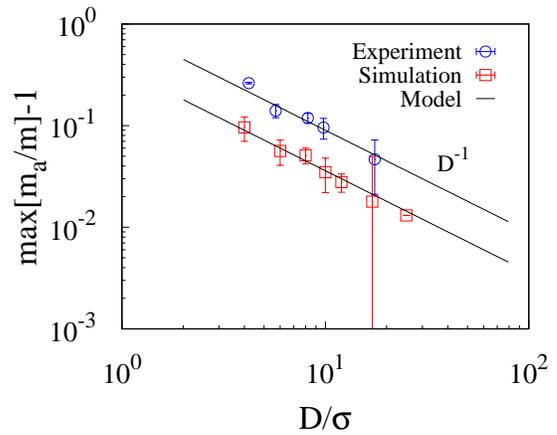}
 \caption{
 Experimental and numerical results for the dependence of the maximum of the reported anomaly on the cylinder diameter. 
\label{fig:model}
}
\end{figure}

According to the model, which is analytically solvable for $h < h^*$, the maximum of the apparent mass scales as 
\begin{equation}
\max\left[\frac{m_a}{m}\right]-1 = \frac{4cn_c(D)}{\pi \rho g D^2}\left(1-\frac{h_a}{h^*}\right) \sim \frac{F_c}{D}.
\label{eq:maxan}
\end{equation}
As we use a constant value of $F_c$ to describe both the experimental and the numerical results, from Eq.~\ref{eq:maxan} we predict that the maximum of the anomaly should scale as $1/D$. We verify this prediction in Fig.~\ref{fig:model}. This scaling further confirms that for large $D$, the anomaly should become indetectable, consistent with our experimental findings.

We have demonstrated via both experiments and simulations a reverse Janssen effect in granular columns, whereby the apparent mass overcomes the true mass.
We note, however, that while the agreement between the two is qualitatively remarkable, the maximum values of $m_a/m$ in the experiment are larger than those in the simulation.
This could arise from the value of the friction coefficient used in the simulation, as discussed in~\cite{SM}, or from the differences in the granular packing preparation protocol; in experiments, the particles are deposited in small chunks while in the simulations they are sequentially added one by one.
Despite this fact, the anomaly clearly results from the existence of compressive frictional forces at the container wall.
We have investigated the dependence of this effect on the various control parameters, and clarified its physical origin.
Finally, we have introduced a model able to describe the observed findings.
This model clarifies that, while the elastic properties of granular systems might be described via a continuum approach, the constitutive relations are generally more complex than previously envisaged~\cite{VanelPRL}: not only do they inherit the history dependence of the frictional interaction, but they also change as the external load varies.

\begin{acknowledgments}
We thank the Singapore Ministry of Education through the Academic Research Fund (Tier 2) MOE2017-T2-1-066 (S), the National Science Foundation (DMR-1609841), the National Research Foundation Singapore and MCIU/AEI/FEDER,UE (PGC2018-097842-B-I00). We are also grateful to the National Supercomputing Centre (NSCC) of Singapore for providing computational resources.
\end{acknowledgments}


\begin{thebibliography}{18}
\expandafter\ifx\csname natexlab\endcsname\relax\def\natexlab#1{#1}\fi
\expandafter\ifx\csname bibnamefont\endcsname\relax
  \def\bibnamefont#1{#1}\fi
\expandafter\ifx\csname bibfnamefont\endcsname\relax
  \def\bibfnamefont#1{#1}\fi
\expandafter\ifx\csname citenamefont\endcsname\relax
  \def\citenamefont#1{#1}\fi
\expandafter\ifx\csname url\endcsname\relax
  \def\url#1{\texttt{#1}}\fi
\expandafter\ifx\csname urlprefix\endcsname\relax\def\urlprefix{URL }\fi
\providecommand{\bibinfo}[2]{#2}
\providecommand{\eprint}[2][]{\url{#2}}

\bibitem[{\citenamefont{Janssen}(1895)}]{Janssen}
\bibinfo{author}{\bibfnamefont{H.~A.} \bibnamefont{Janssen}},
  \bibinfo{journal}{Z. Verein Deutsch. Ing.} \textbf{\bibinfo{volume}{39}},
  \bibinfo{pages}{1045} (\bibinfo{year}{1895}).

\bibitem[{\citenamefont{Sperl}(2006)}]{Sperl}
\bibinfo{author}{\bibfnamefont{M.}~\bibnamefont{Sperl}},
  \bibinfo{journal}{Granular Matter} \textbf{\bibinfo{volume}{8}},
  \bibinfo{pages}{59} (\bibinfo{year}{2006}).

\bibitem[{\citenamefont{Jaeger et~al.}(1996)\citenamefont{Jaeger, Nagel, and
  Behringer}}]{RevJaeger}
\bibinfo{author}{\bibfnamefont{H.~M.} \bibnamefont{Jaeger}},
  \bibinfo{author}{\bibfnamefont{S.~R.} \bibnamefont{Nagel}}, \bibnamefont{and}
  \bibinfo{author}{\bibfnamefont{R.~P.} \bibnamefont{Behringer}},
  \bibinfo{journal}{Rev. Mod. Phys.} \textbf{\bibinfo{volume}{68}},
  \bibinfo{pages}{1259} (\bibinfo{year}{1996}).

\bibitem[{\citenamefont{Andreotti et~al.}(2013)\citenamefont{Andreotti,
  Forterre, and Pouliquen}}]{Andreotti}
\bibinfo{author}{\bibfnamefont{B.}~\bibnamefont{Andreotti}},
  \bibinfo{author}{\bibfnamefont{Y.}~\bibnamefont{Forterre}}, \bibnamefont{and}
  \bibinfo{author}{\bibfnamefont{O.}~\bibnamefont{Pouliquen}},
  \bibinfo{journal}{Cambridge University Press}  (\bibinfo{year}{2013}).

\bibitem[{\citenamefont{Nedderman}(1992)}]{Nedderman1992}
\bibinfo{author}{\bibfnamefont{R.~M.} \bibnamefont{Nedderman}},
  \emph{\bibinfo{title}{{Statics and Kinematics of Granular Materials}}}
  (\bibinfo{publisher}{Cambridge University Press},
  \bibinfo{address}{Cambridge}, \bibinfo{year}{1992}), ISBN
  \bibinfo{isbn}{9780511600043}.

\bibitem[{\citenamefont{Vanel and Cl\'{e}ment}(1999)}]{VanelEPJB}
\bibinfo{author}{\bibfnamefont{L.}~\bibnamefont{Vanel}} \bibnamefont{and}
  \bibinfo{author}{\bibfnamefont{E.}~\bibnamefont{Cl\'{e}ment}},
  \bibinfo{journal}{E. Eur. Phys. J. B} \textbf{\bibinfo{volume}{11}},
  \bibinfo{pages}{525} (\bibinfo{year}{1999}).

\bibitem[{\citenamefont{Vanel et~al.}(2000)\citenamefont{Vanel, Claudin,
  Bouchaud, Cates, Cl\'ement, and Wittmer}}]{VanelPRL}
\bibinfo{author}{\bibfnamefont{L.}~\bibnamefont{Vanel}},
  \bibinfo{author}{\bibfnamefont{P.}~\bibnamefont{Claudin}},
  \bibinfo{author}{\bibfnamefont{J.-P.} \bibnamefont{Bouchaud}},
  \bibinfo{author}{\bibfnamefont{M.~E.} \bibnamefont{Cates}},
  \bibinfo{author}{\bibfnamefont{E.}~\bibnamefont{Cl\'ement}},
  \bibnamefont{and} \bibinfo{author}{\bibfnamefont{J.~P.}
  \bibnamefont{Wittmer}}, \bibinfo{journal}{Phys. Rev. Lett.}
  \textbf{\bibinfo{volume}{84}}, \bibinfo{pages}{1439} (\bibinfo{year}{2000}).

\bibitem[{\citenamefont{Bratberg et~al.}(2005)\citenamefont{Bratberg, Maloy,
  and Hansen}}]{BratbergUpward}
\bibinfo{author}{\bibfnamefont{I.}~\bibnamefont{Bratberg}},
  \bibinfo{author}{\bibfnamefont{K.}~\bibnamefont{Maloy}}, \bibnamefont{and}
  \bibinfo{author}{\bibfnamefont{A.}~\bibnamefont{Hansen}},
  \bibinfo{journal}{Eur. Phys. J. E} \textbf{\bibinfo{volume}{18}},
  \bibinfo{pages}{245} (\bibinfo{year}{2005}).

\bibitem[{\citenamefont{Shaxby and Evans}(1923)}]{Shaxby1923}
\bibinfo{author}{\bibfnamefont{J.~H.} \bibnamefont{Shaxby}} \bibnamefont{and}
  \bibinfo{author}{\bibfnamefont{J.~C.} \bibnamefont{Evans}},
  \bibinfo{journal}{Trans. Faraday Soc} \textbf{\bibinfo{volume}{19}},
  \bibinfo{pages}{60} (\bibinfo{year}{1923}).

\bibitem[{\citenamefont{Wittmer et~al.}(1997)\citenamefont{Wittmer, Cates, and
  Claudin}}]{Wittmer1997}
\bibinfo{author}{\bibfnamefont{J.~P.} \bibnamefont{Wittmer}},
  \bibinfo{author}{\bibfnamefont{M.~E.} \bibnamefont{Cates}}, \bibnamefont{and}
  \bibinfo{author}{\bibfnamefont{P.}~\bibnamefont{Claudin}},
  \bibinfo{journal}{Journal de Physique I} \textbf{\bibinfo{volume}{7}},
  \bibinfo{pages}{39} (\bibinfo{year}{1997}), ISSN \bibinfo{issn}{1155-4304}.

\bibitem[{\citenamefont{Jotaki and
  Moriyama}(1977)}]{JotakiDependFillingProcedure}
\bibinfo{author}{\bibfnamefont{T.}~\bibnamefont{Jotaki}} \bibnamefont{and}
  \bibinfo{author}{\bibfnamefont{R.}~\bibnamefont{Moriyama}},
  \bibinfo{journal}{Journal of the Research Association of Powder Technology,
  Japan} \textbf{\bibinfo{volume}{14}}, \bibinfo{pages}{609}
  (\bibinfo{year}{1977}).

\bibitem[{\citenamefont{Wittmer et~al.}(1996)\citenamefont{Wittmer, Claudin,
  Cates, and Bouchaud}}]{WittmerLetterstoNature}
\bibinfo{author}{\bibfnamefont{J.~P.} \bibnamefont{Wittmer}},
  \bibinfo{author}{\bibfnamefont{P.}~\bibnamefont{Claudin}},
  \bibinfo{author}{\bibfnamefont{M.~E.} \bibnamefont{Cates}}, \bibnamefont{and}
  \bibinfo{author}{\bibfnamefont{J.~P.} \bibnamefont{Bouchaud}},
  \bibinfo{journal}{Nature} \textbf{\bibinfo{volume}{382}},
  \bibinfo{pages}{336} (\bibinfo{year}{1996}).

\bibitem[{\citenamefont{Plimpton}(1995)}]{Plimpton1995a}
\bibinfo{author}{\bibfnamefont{S.~J.} \bibnamefont{Plimpton}},
  \bibinfo{journal}{J. Comput. Phys.} \textbf{\bibinfo{volume}{117}},
  \bibinfo{pages}{1} (\bibinfo{year}{1995}).

\bibitem[{\citenamefont{Silbert et~al.}(2001)\citenamefont{Silbert,
  Erta\ifmmode~\mbox{\c{s}}\else \c{s}\fi{}, Grest, Halsey, Levine, and
  Plimpton}}]{Silbert}
\bibinfo{author}{\bibfnamefont{L.~E.} \bibnamefont{Silbert}},
  \bibinfo{author}{\bibfnamefont{D.}~\bibnamefont{Erta\ifmmode~\mbox{\c{s}}\else
  \c{s}\fi{}}}, \bibinfo{author}{\bibfnamefont{G.~S.} \bibnamefont{Grest}},
  \bibinfo{author}{\bibfnamefont{T.~C.} \bibnamefont{Halsey}},
  \bibinfo{author}{\bibfnamefont{D.}~\bibnamefont{Levine}}, \bibnamefont{and}
  \bibinfo{author}{\bibfnamefont{S.~J.} \bibnamefont{Plimpton}},
  \bibinfo{journal}{Phys. Rev. E} \textbf{\bibinfo{volume}{64}},
  \bibinfo{pages}{051302} (\bibinfo{year}{2001}).

\bibitem[{SM()}]{SM}
\bibinfo{note}{See Supplemental Material at
  XXX for details
  on the numerical model, on the spatially resolved frictional forces acting on
  the wall, the role of friction, the scaling of the number of particle-wall
  contacts with the cylinder diameter.}

\bibitem[{\citenamefont{Mindlin}(1949)}]{49Min}
\bibinfo{author}{\bibfnamefont{R.}~\bibnamefont{Mindlin}},
  \bibinfo{journal}{Trans. ASME} \textbf{\bibinfo{volume}{16}},
  \bibinfo{pages}{259} (\bibinfo{year}{1949}).

\bibitem[{\citenamefont{Hentschel et~al.}(2017)\citenamefont{Hentschel,
  Jaiswal, Mondal, Procaccia, and Zylberg}}]{Hentschel}
\bibinfo{author}{\bibfnamefont{H.~G.~E.} \bibnamefont{Hentschel}},
  \bibinfo{author}{\bibfnamefont{P.~K.} \bibnamefont{Jaiswal}},
  \bibinfo{author}{\bibfnamefont{C.}~\bibnamefont{Mondal}},
  \bibinfo{author}{\bibfnamefont{I.}~\bibnamefont{Procaccia}},
  \bibnamefont{and} \bibinfo{author}{\bibfnamefont{J.}~\bibnamefont{Zylberg}},
  \bibinfo{journal}{Soft Matter} \textbf{\bibinfo{volume}{13}},
  \bibinfo{pages}{5008} (\bibinfo{year}{2017}).

\bibitem[{\citenamefont{Condre et~al.}(2007)\citenamefont{Condre, Ligoure, and
  Cipelletti}}]{Condre2007}
\bibinfo{author}{\bibfnamefont{J.-M.} \bibnamefont{Condre}},
  \bibinfo{author}{\bibfnamefont{C.}~\bibnamefont{Ligoure}}, \bibnamefont{and}
  \bibinfo{author}{\bibfnamefont{L.}~\bibnamefont{Cipelletti}},
  \bibinfo{journal}{Journal of Statistical Mechanics: Theory and Experiment}
  \textbf{\bibinfo{volume}{2007}}, \bibinfo{pages}{P02010}
  (\bibinfo{year}{2007}).

\end{thebibliography}

\pagebreak
~\newpage
\onecolumngrid
\renewcommand*{\thefigure}{S\arabic{figure}}
\setcounter{figure}{0} 
\section{Numerical details}
In the numerical simulations, the interaction between two particles has a normal and a tangential component. The tangential forces lead to a torque on the particles, driving their rotational motion. 
We model the interaction forces using the standard linear spring-dashpot model.
Specifically, we use the model L3 described in L. E. Silbert, D. Ertas, G. S. Grest, T. C. Halsey, D. Levine, and S. J. Plimpton, Phys. Rev. E. {\bf 64}, 051302 (2001), also adopting the values of the parameters suggested there. 
We briefly recap the model below.

The normal component is given by:
\begin{equation}
    {\bf f}_n = k_n \delta_{ij} {\bf n} - M_{\rm eff}\gamma_n v_n {\bf n}
\end{equation}
where $\delta_{ij} = (\sigma_i+\sigma_j)/2-r_{ij}$ is the particle overlap, $\sigma_i$ and $\sigma_j$ the diameters of the particles, ${\bf r}_{ij} = {\bf r}_{i} - {\bf r}_{j}$, ${\bf n} = {\bf r}_{ij}/|{\bf r}_{ij}|$, and 
$M_{\rm eff} = M_i M_j/(M_i+M_j)$ is the effective mass of the interacting particles.
The force is only active when the particles overlap, $\delta_{ij} > 0$.
The constants $\gamma_n$ and $k_n$ are the normal damping coefficient and elastic constant, respectively, and $v_n = ({\bf v}_j - {\bf v}_i) \cdot {\bf n}$, with ${\bf v_i}$ the velocity of particle $i$ during the rearrangement.

To model the tangential interaction, we keep track of the tangential elastic displacement
$\bm{\xi}_t = \int_{t_0}^t  \bm{v}_t dt$, the integral of the relative tangential velocity, ${\bf v}_t = [({\bf v}_j - {\bf v}_i) \cdot {\bf t}] {\bf t}$, at the contact point over the duration of the contact.
Note that ${\bf v}_t$ depends on both the translational and the rotational motion of the particles.
Care is taken to ensure that $\bm{\xi}_t$ is always orthogonal to $\bf n$. The tangential force is then the sum of an elastic and a viscous damping term, 
\begin{equation}
   \mathbf{f}_t =  -k_t \bm{\xi} - M_{\rm eff} \gamma_t {\bf v}_t,
\end{equation}
with $k_t$ an elastic constant and
$\gamma_t$ a tangential damping coefficient.
The Coulomb condition is enforced by rescaling the magnitude of $\bm{\xi}$ to ensure that 
$|{\bf f}_t| \leq \mu |{\bf f}_n|$, where $\mu$ is the friction coefficient.

The particle wall interaction is described as the interaction between two particles, one of which has infinite radius and mass.

The numerical results in the main text are for a value of the friction coefficient $\mu = 0.2$, but we have considered other values as described in the following.
Simulations are performed using the LAMMPS software package, S. J. Plimpton, J. Comput. Phys. {\bf 117}, (1995).

\newpage
\section{Frictional forces acting on the wall}
\begin{figure}[t!]
 \includegraphics[width=0.48\textwidth]{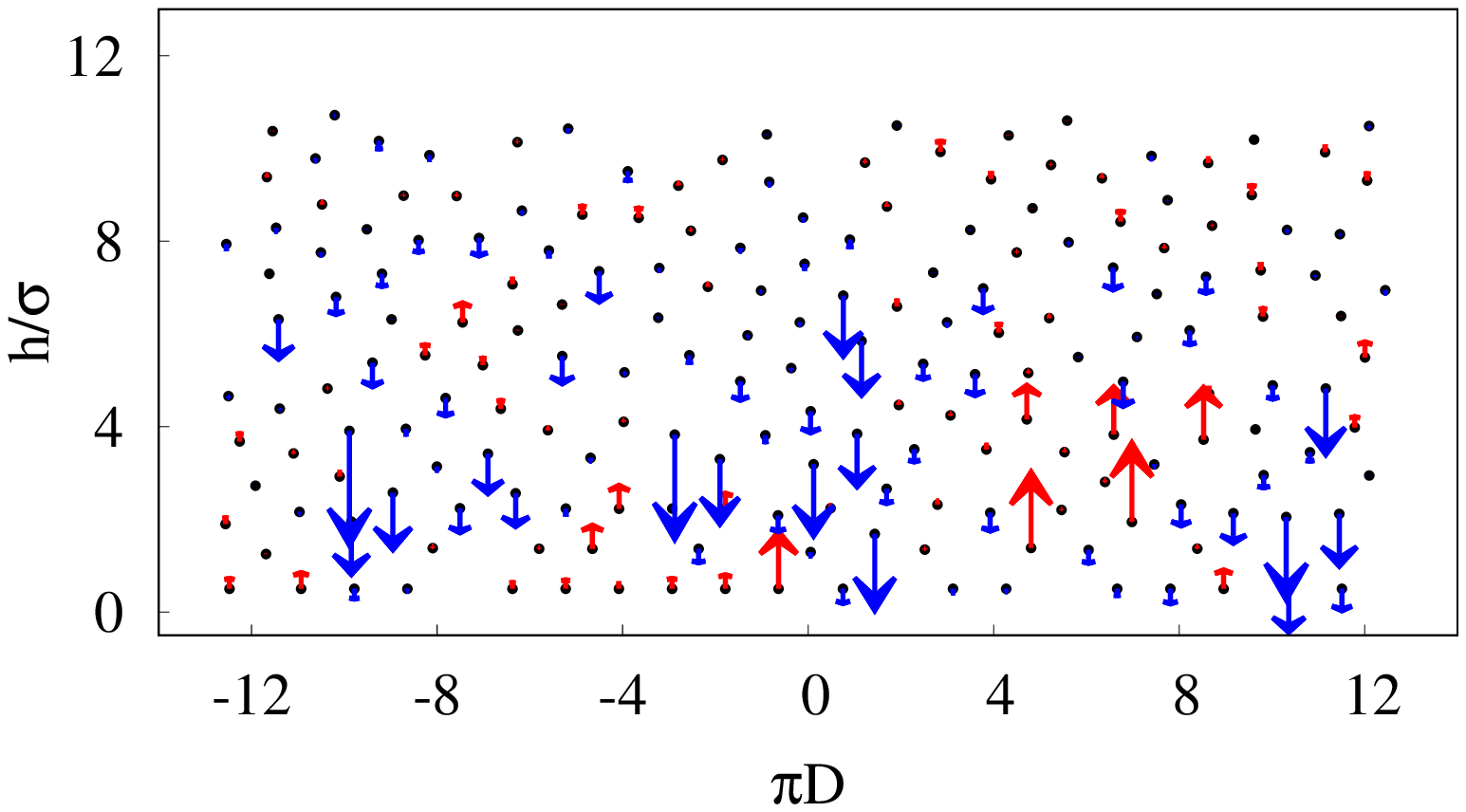}
 \caption{
 Vertical component of the wall-particle frictional forces, for cylinder with diameter $D=8 \sigma$. The friction coefficient is $\mu=0.2$.
\label{fig:anomReg}
}
\end{figure}
In the main text, we have presented data for the average value of the frictional force the container exerts on a particle, as a function of the depth of the particle.
Here we resolve the single particle forces.
We unwrap the cylinder of diameter $D$ into a rectangle with length $l = \pi D$.
We then associate a point ($l_i,h_i$) within the rectangle to the contact between the wall and particle $i$ in position $(x_i, y_i, h_i)$.
Here $l_i = (D/2) \psi_i$, with $-\pi < \psi_i \leq \pi$ the polar angle of particle $i$. 
Fig.~\ref{fig:anomReg} illustrates the cylinder-particle contacts.

We represent the frictional forces associated to these contacts via arrows, whose length is proportional to the magnitude of the force.
Red arrows correspond to supportive forces, while blue arrows correspond to compressive forces. 
For a cylinder with diameter $D = 8 \sigma$ filled up to  $\simeq 11 \sigma$, with $\mu = 0.2$, in a given configuration, we clearly observe that the majority of the forces are compressive. 

We similarly observe a prevalence of compressive forces close to the free surface of filled cylinders, regardless of their filling height.

\section{Role of friction}
\begin{figure}[tb]
 \centering
 \includegraphics[width=0.35\textwidth]{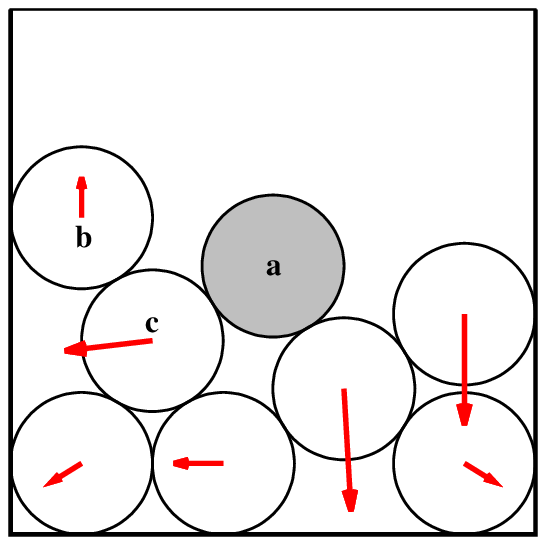}
 \caption{
 Fig 4b of the main text. Arrows are proportional to the displacement induced by the addition of grain ${\bf a}$ to the column.
\label{fig:wedge}
}
\end{figure}

In our interpretation, summarized in Fig. 4b of the main text and reproduced in Fig.~\ref{fig:wedge}, compressive forces originate as the addition of a grain, {\bf a} in the figure, induces the motion of other grains in contact with the wall, {\bf b} in the figure.  
This occurs as ${\bf a}$ pushes radially outwards {\bf c}, which in turn pushes upwards grain ${\bf b}$ by acting as a wedge. 
This upward motion leads to a growth of $\mathbf{\xi}_t$, the relative particle-wall shear displacement at the point of contact.
In the presence of friction, the motion of particle ${\bf c}$ also induces a torque on particle ${\bf b}$ and its clockwise rotation. This also contributes to $\mathbf{\xi}_t$, and we have numerically verified that most often the rotational contribution is dominant.
The growth of $\mathbf{\xi}_t$ leads to the emergence of compressive frictional forces.

\begin{figure}[!t]
 \centering
 \includegraphics[angle=0,width=0.45\textwidth]{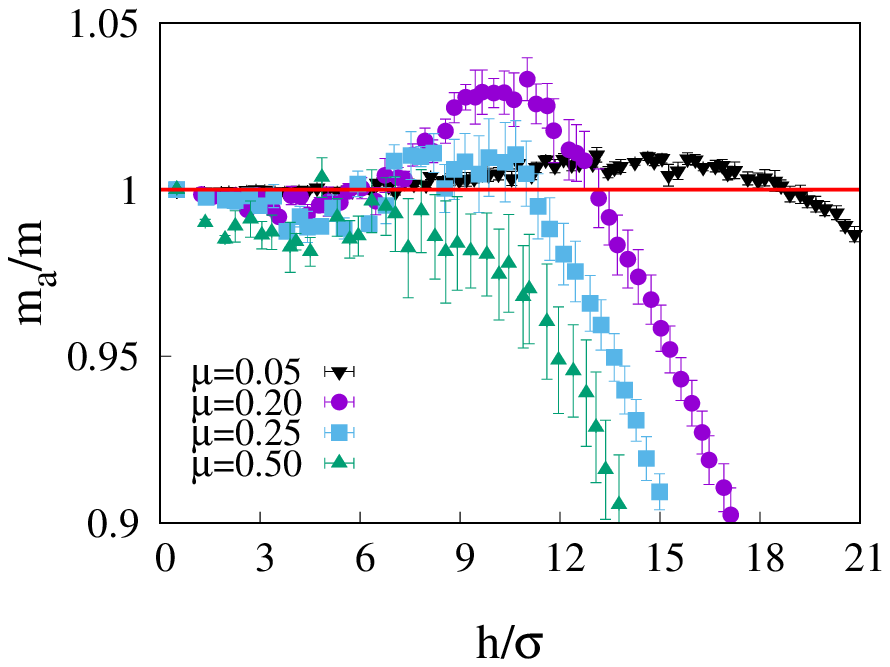}
 \caption{
Dependence of the ratio $m_a/m$ on the filling height, in numerical simulations conducted in a cylinder with diameter $D = 8\sigma$, for different values of the friction coefficient.
\label{fig:friction}
}
\end{figure}

The effect of friction in this process is understood in analogy with what has been observed in the context of granular piles. 
In that case, the friction coefficient sets the angle of repose of the pile, which grows with the friction coefficient. See, e.g. Hentschel, H. G. E., Jaiswal, P. K., Mondal, C., Procaccia, I., \& Zylberg, J., Soft Matter, {\bf 13}, 5008 (2017).
Accordingly, the higher the friction coefficient the lower the tendency of the particles to move radially outwards. 
Since the radial outward movement of the particles is involved in the generation of the compressive forces, we expect these compressive forces to be suppressed in the limit of high friction.
Hence, in this limit no anomaly should be observed.
In the opposite limit of small friction there will be large radial displacements, and hence upward vertical displacements of particles in contact with the wall.
However, in this case, since $\mu$ is small, there are essentially no frictional forces at the wall, and hence again no anomaly should be observed.
Since we expect the anomaly to vanish both in the limit of small and high friction, we predict a non-monotonic dependence of the magnitude of the associated effect on the friction coefficient.
We have found that this is indeed the case.
As an example, we illustrate in Fig.~\ref{fig:friction} numerical results for the dependence of $m_a/m$ on the relative filling height $h/\sigma$ for different values of $\mu$. 
The non-monotonic dependence of the strength of the anomaly on the friction coefficient is apparent.


\section{Number of particle-wall contacts}
\begin{figure}[t!]
 \centering
 \includegraphics[angle=0,width=0.48\textwidth]{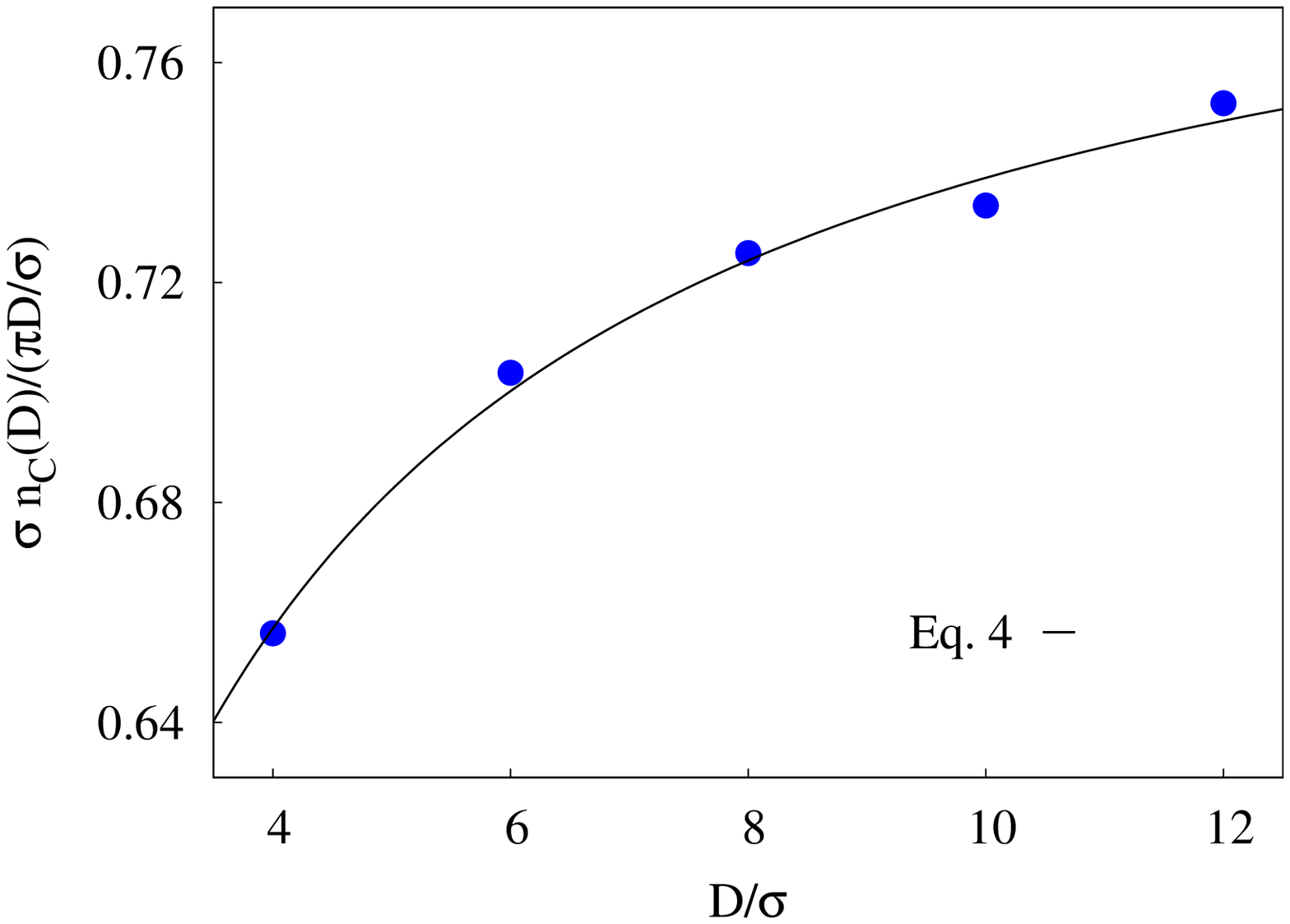}
 \caption{
  Diameter dependence of the number of particles in contact with wall of a cylinder, per unit length.  The full line is a fit to Eq.~\ref{eq:ncd}.
\label{fig:nc}
}
\end{figure}
The shear stress the wall exerts at depth $z$ is 
\begin{equation}
\tau(z) \:=\: \frac{\langle F_w(z) \rangle n_c(D)}{\pi D}
\end{equation}
where $\langle F_w(z) \rangle$ is the average frictional force the wall exerts on a particle at depth $z$, and $n_c(D)$ is the number of particle-wall contacts per unit cylinder length.
The shear stress is responsible for a normal force on the grains of magnitude $\tau(z) \pi D/2$, per unit length; hence, its contribution to the pressure gradient is $\frac{4}{D}\tau(z)$, as in Eq.~1 in the main text.
To evaluate the pressure dependence on the depth we therefore need a model for $n_c(D)$.

To set-up this model we consider that, in disordered packing, particles have positions correlated over a length scale $\xi_c$ of a few diameters; only at larger distances, the radial distribution function is $g(r) = 1$. 
The precise value depends on the preparation protocol.
The presence of this length scale affects the scaling of the number of particles in contact with the cylinder wall.
In particular, for $D \gg \xi_c$ the number of particle-cylinder contacts, per unit-length of the cylinder, is proportional to the cylinder circumference: $n_c(D) = \frac{a\pi D}{\sigma}$, with $a < 1$. 
The actual number of contacts is suppressed with respect to the above prediction when $D < \xi_c$. 
We have numerically investigated the dependence of $n_c$ on the cylinder diameter, and find that this relation is well described by
\begin{equation}
\sigma n_c(D)  = a\pi \left(\frac{D}{\sigma}\right) \frac{1}{1+b\sigma/D},  
\label{eq:ncd}
\end{equation}
where $a\simeq 0.8$ and $b\simeq0.9$ are fitting parameters; see Fig.~\ref{fig:nc}. 
We have used the above relationship in the theoretical model for the evaluation of the stress discussed in the main text.\\
\section{Disappearance of the anomaly is narrow cylinders}
\begin{figure}[!b]
 \centering
 \includegraphics[angle=0,width=0.48\textwidth]{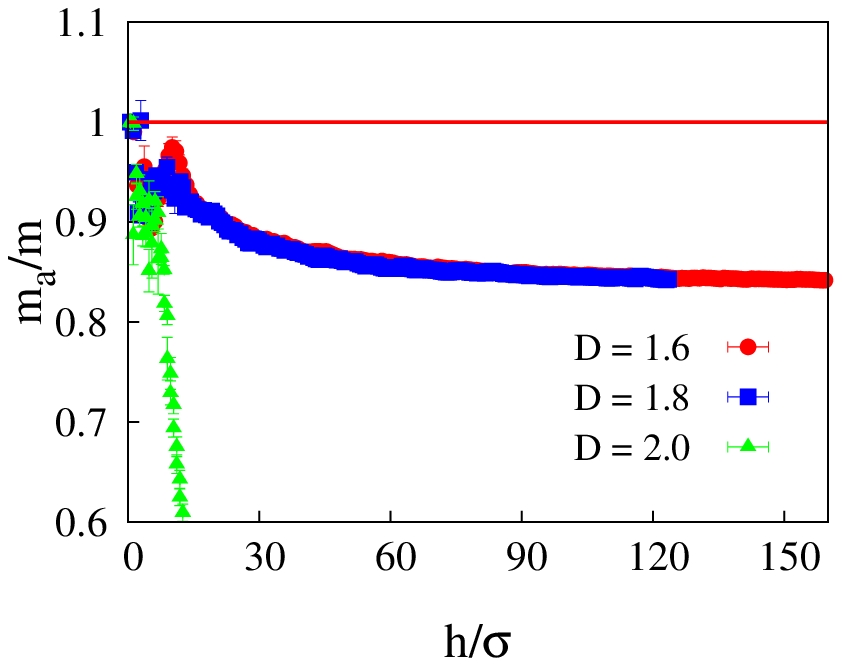}
 \caption{
  Disappearance of reverse Janssen effect for very narrow cylinder with $D < 2\sigma$.
\label{fig:disappearSmall}
}
\end{figure}
The reverse Jannsen effect described in our manuscript disappears as $D$ increases. 
Here we demonstrate that the effect is also suppressed in very narrow cylinders, that is, for $D < 2\sigma$.
We show this is the case in Fig.~\ref{fig:disappearSmall}, where we plot the ratio between apparent and added mass, $m_a/m$, as a function of the filling height $h$ normalized by the particle diameter $\sigma$, for cylinders with very small $D$; we find $m_a < m$ for all values of $h/\sigma$.

\end{document}